\documentclass[preprint,showpacs,nofootinbib,preprintnumbers,amsmath,amssymb,epsfig]{revtex4}

\usepackage{graphicx}
\usepackage{bm}
\usepackage{lscape}

\begin{document}


\title{Direct $CP$ violation in $\bar{B}^0 \rightarrow\rho^0(\omega)\rho^0(\omega)\rightarrow\pi^+\pi^-\pi^+\pi^-$}

\author{X.-H. Guo\footnote{email: xhguo@bnu.edu.cn}}
\author{Z.-H. Zhang\footnote{email: zhangzh@brc.bnu.edu.cn}}%
\affiliation{Institute of Low Energy Nuclear Physics, Beijing
Normal University, Beijing 100875, China.\\}


\begin{abstract}

We study the direct $CP$ violation in
$\bar{B}^0\rightarrow\rho^0(\omega)\rho^0(\omega)\rightarrow\pi^+\pi^-\pi^+\pi^-$
(with unpolarized $\rho^0(\omega)$) via the $\rho-\omega$ mixing
mechanism which causes a large strong phase difference and
consequently a large $CP$ violating asymmetry when the masses of
the $\pi^+\pi^-$ pairs are in the vicinity of the $\omega$
resonance. Since there are two $\rho (\omega)$ mesons in the
intermediate state $\rho-\omega$ mixing contributes twice to the
first order of isospin violation, leading to an even larger $CP$
violating asymmetry (could be 30\% -- 50\% larger) than in the
case where only one $\rho (\omega)$ meson is involved. The $CP$
violating asymmetry depends on the Cabibbo-Kobayashi-Maskawa (CKM)
matrix elements and the hadronic matrix elements. The
factorization approach is applied in the calculation of the
hadronic matrix elements with the nonfactorizable effects being
included effectively in an effective parameter, $N_c$. We give the
constraint on the range of $N_c$ from the latest experimental data
for the branching ratios for $\bar{B}^0 \rightarrow\rho^0\rho^0$
and $\bar{B}^0 \rightarrow\rho^+\rho^-$. We find that the $CP$
violating asymmetry could be very large (even more than 90\% for
some values of $N_c$). It is shown that the sensitivity of the
$CP$ violating asymmetry to $N_c$ is large compared with its
smaller sensitivity to the CKM matrix elements. We also discuss
the possibility to remove the mod $(\pi)$ ambiguity in the
determination of the $CP$ violating phase angle $\alpha$ through
the measurement of the $CP$ violating asymmetry in the decay
$\bar{B}^0\rightarrow\rho^0(\omega)\rho^0(\omega)\rightarrow\pi^+\pi^-\pi^+\pi^-$.
\end{abstract}


\maketitle

\section{\label{intro}Introduction}

Although $CP$ violation has been a central concern in particle
physics since it was first observed in the neutral kaon system
more than four decades ago \cite{christenson} the dynamical origin
of $CP$ violation still remains an open problem. $CP$ violation in
the framework of the Standard Model (SM) is supposed to arise from
a weak complex phase in the Cabibbo-Kobayashi-Maskawa (CKM) matrix
which is based on quark flavor mixing \cite{cab,kob}. Therefore,
the study of $CP$ violation is essential to the test of the CKM
mechanism in the SM.

Besides the kaon system much more studies have been carried out on
$CP$ violation in the $B$ meson system both theoretically and
experimentally in the past few years. It was suggested
theoretically that large $CP$ violating asymmetries should be
observed in the experiments for $B$ mesons \cite{carter}. This
important prediction has already been confirmed by the experiments
of BaBar and Belle etc. through the measurements on $CP$ violation
in several decay channels of $B$ mesons such as $B^0 \rightarrow
J/\psi K_S^0$ and $B^0 \rightarrow  K^+ \pi^-$ \cite{cpexp}. From
the summer of 2007, the Large Hadron Collider (LHC) at CERN will
start to contribute to the exploration of $CP$ violation in the
$B$ meson system in a more accurate way due to its much higher
statistics. This will also provide an opportunity to discover new
physics beyond the SM.

In the decay process we have the so-called direct $CP$ violation
which occurs through the interference of two amplitudes with
different weak phases and strong phases. The weak phase difference
is directly determined by the CKM matrix. On the contrary, the
strong phase is usually due to complicated strong interaction and
hence difficult to control. Since a large strong phase difference
is required for a large $CP$ asymmetry, one needs to appeal to
some phenomenological mechanism to get such a large strong phase
difference. The charge asymmetry violating mixing between $\rho^0$
and $\omega$ ($\rho-\omega$ mixing) has been applied for this
purpose in the past few years. From a series of studies for $CP$
violation in some decay channels of heavy hadrons including $B$,
$\Lambda_b$ and $D$, it has been found that $\rho-\omega$ mixing
can provide a very large strong phase difference (usually 90
degrees) when the mass of the decay product of $\rho$($\omega$),
$\pi^+ \pi^-$, is in the vicinity of the $\omega$ resonance
\cite{eno, gar, guo1, guo2, lei}. Furthermore, it has been shown
that the measurement of the $CP$ violating asymmetry for these
decays can be used to remove the mod $(\pi)$ ambiguity in the
determination of the $CP$ violating phase angle $\alpha$.

In this paper, we will investigate the $CP$ violating asymmetry
for the decay
$\bar{B}^0\rightarrow\rho^0(\omega)\rho^0(\omega)\rightarrow\pi^+\pi^-\pi^+\pi^-$.
This process is unique since it has two $\rho (\omega)$ mesons in
the intermediate state, each of them contributing $\rho-\omega$
mixing. One can expect that there should be a bigger $CP$
violating asymmetry than in the case where $\rho-\omega$ mixing
only contributes once. It will be shown from our explicit
calculations that this is true indeed. The $CP$ violating
asymmetry in the case of double $\rho-\omega$ mixing could be 30
-- 50\% bigger than that in the case of single $\rho-\omega$
mixing, depending on the value of $N_c$ and $q^2/m_b^2$ (see the
meaning of $N_c$ and $q^2/m_b^2$ below).

In our calculations of the $CP$ violating asymmetry, hadronic
matrix elements for both tree and penguin operators in the
effective Hamiltonian are involved. These matrix elements are
controlled by the effects of nonperturbative QCD which are
difficult to handle. In order to extract the strong phase
difference we will use the factorization approximation, in which
one of the currents in the Hamiltonian is factorized out and
generates a meson, assuming the vacuum intermediate state
saturation.  In this way, the decay amplitude becomes the product
of two matrix elements. Such factorization scheme was first argued
to be plausible in energetic decays like bottom-hadron decays
\cite{bjorken}\cite {dugan}, then was proved to be the leading
order result in the framework of QCD factorization when the
radiative QCD corrections of order $O(\alpha_s (m_b))$ ($m_b$ is
the b-quark mass) and the $O(1/m_b)$ corrections in the heavy
quark effective theory are neglected \cite{qcdf}. Since the
nonfactorizable contributions are ignored in the factorization
scheme we introduce an effective parameter, $N_c$, in order to
take into account nonfactorizable contributions effectively. In
this way, the value of $N_c$ is not the color number (3) any more,
but should be determined by experimental data. In the present
work, this will be done by comparing the theoretical results with
the experimental data for the decay branching ratios for the
processes $\bar{B}^0 \rightarrow\rho^0\rho^0$ and $\bar{B}^0
\rightarrow\rho^+\rho^-$.

The remainder of this paper is organized as follows. In Sec.
\ref{sec:hamckm} we briefly present the effective Hamiltonian, the
Wilson coefficients and the CKM matrix elments. In Sec.
\ref{sec:cpv1} we give the formalism for the $CP$ violating
asymmetry in $\bar{B}^0\rightarrow\rho^0 (\omega) \rho^0 (\omega)
\rightarrow \pi^+\pi^-\pi^+\pi^-$ via $\rho-\omega$ mixing. Then
we give the calculation details of the strong phase difference and
the numerical results for the $CP$ violating asymmetry. In Sec.
\ref{sec:branchratio}, we calculate the branching ratios for
$\bar{B}^0\rightarrow\rho^0\rho^0$ and $\bar{B}^0
\rightarrow\rho^+\rho^-$ and present the range of $N_c$ allowed by
the latest experimental data for these decays. In the last
section, we give a summary and discussion.

\section{\label{sec:hamckm}The effective
hamiltonian and the CKM matrix}

In order to calculate the direct $CP$ violating asymmetry one
needs to use the following effective weak Hamiltonian based on the
operator product expansion \cite{buc}:
\begin{eqnarray}
H_{\Delta B=1}=\frac{G_F}{\sqrt[]{2}}[\sum_{q=d,s}V_{ub} V_{uq}^*
(c_1 O_1^u +c_2 O_2^u)-V_{tb} V_{tq}^*\sum_{i=3}^{10}c_i O_i]+H.
c., \label{ham}
\end{eqnarray}
where $c_i$(i=1,...,10) are the Wilson coefficients, $V_{ub}$,
$V_{uq}$, $V_{tb}$ and $V_{tq}$ are the CKM matrix elements. The
operators $O_i$ have the following form:
\begin{eqnarray}
&&\ O_1^u =\bar{q}_\alpha \gamma_\mu(1-\gamma_5)u_\beta
\bar{u}_\beta\gamma^\mu(1-\gamma_5)b_\alpha,\nonumber\\
&&\ O_2^u
=\bar{q}\gamma_\mu(1-\gamma_5)u\bar{u}\gamma^\mu(1-\gamma_5)b,\nonumber\\
&&\ O_3
=\bar{q}\gamma_\mu(1-\gamma_5)b\sum_{q^\prime}\bar{q^\prime}\gamma^\mu(1-\gamma_5)q^\prime,\nonumber\\
&&\ O_4
=\bar{q}_\alpha \gamma_\mu(1-\gamma_5)b_\beta\sum_{q^\prime}\bar{q^\prime}_\beta\gamma^\mu(1-\gamma_5)q_\alpha^\prime,\nonumber\\
&&\ O_5
=\bar{q}\gamma_\mu(1-\gamma_5)b\sum_{q^\prime}\bar{q^\prime}\gamma^\mu(1+\gamma_5)q^\prime,\nonumber\\
&&\ O_6
=\bar{q}_\alpha \gamma_\mu(1-\gamma_5)b_\beta\sum_{q^\prime}\bar{q^\prime}_\beta\gamma^\mu(1+\gamma_5)q_\alpha^\prime,\nonumber\\
&&\ O_7
=\frac{3}{2}\bar{q}\gamma_\mu(1-\gamma_5)b\sum_{q^\prime}e_{q^\prime}\bar{q^\prime}\gamma^\mu(1+\gamma_5)q^\prime,\nonumber\\
&&\ O_8
=\frac{3}{2}\bar{q}_\alpha\gamma_\mu(1-\gamma_5)b_\beta\sum_{q^\prime}e_{q^\prime}\bar{q^\prime}_\beta\gamma^\mu(1+\gamma_5)q_\alpha^\prime,\nonumber\\
&&\ O_9
=\frac{3}{2}\bar{q}\gamma_\mu(1-\gamma_5)b\sum_{q^\prime}e_{q^\prime}\bar{q^\prime}\gamma^\mu(1-\gamma_5)q^\prime,\nonumber\\
&&\ O_{10}
=\frac{3}{2}\bar{q}_\alpha\gamma_\mu(1-\gamma_5)b_\beta\sum_{q^\prime}e_{q^\prime}\bar{q^\prime}_\beta\gamma^\mu(1-\gamma_5)q_\alpha^\prime,
\label{oper}
\end{eqnarray}
where $\alpha$ and $\beta$ are color indices, and $q^\prime=u, d$
or $s$ quarks. In Eq. (\ref{oper}) $O_1^u$ and $O_2^u$ are tree
operators, $O_3$--$O_6$ are QCD penguin operators, and
$O_7$--$O_{10}$ arise from electroweak penguin diagrams.

The Wilson coefficients are known to the next-to-leading
logarithmic order \cite{buc, buras}. They are renormalization
scheme dependent since the renormalization prescription involves
an arbitrariness in the finite parts in the renormalization
procedure. The physical quantities should be renormalization
scheme independent. Since the radiative QCD corrections are not
included in the factorization approach we work, the hadronic
matrix elements do not carry any information about the
renormalization scheme dependence\footnote{It has been shown that
in the QCD factoization approach the renormalization scheme
dependence of the Wilson coefficients and that of the hadronic
matrix elements cancel \cite{qcdf}.}. Therefore, we choose to use
the renormalization scheme independent Wilson coefficients which
are defined in Refs. \cite{buras, des, fle} so that the $CP$
violating asymmetry we obtain is renormalization scheme
independent. The renormalization scale $\mu$ is chosen as the
energy scale in the decays of the $B$ meson, $O(m_b)$. When
$\mu=5$ GeV, these renormalization scheme independent Wilson
coefficients take the following values \cite{des,fle}:
\begin{eqnarray}
&&\ c_1=-0.3125, \quad c_2=1.1502,\nonumber\\ &&\ c_3=0.0174,\quad
c_4=-0.0373,\nonumber\\ &&\ c_5=0.0104,\quad c_6=-0.0459,\nonumber\\
&&\ c_7=-1.050\times 10^{-5},\quad c_8=3.839\times 10^{-4},\nonumber\\
&&\ c_9=-0.0101,\quad c_{10}=1.959\times 10^{-3}. \label{cvalue}
\end{eqnarray}

The matrix elements of the operators $O_i$ should be renormalized
to the one-loop order. This results in the effective Wilson
coefficients, $c_i^\prime$, which satisfy the constraint
\begin{eqnarray}
c_i(m_b)\langle O_i(m_b)\rangle=c_i^\prime\langle O_i\rangle
^{\rm{tree} }, \label{c}
\end{eqnarray}
where $\langle O_i\rangle ^{\rm{tree}}$ are the matrix elements at
the tree level, which will be evaluated in the factorization
approach. From Eq. (\ref{c}), the relations between $c_i^\prime$
and $c_i$ are \cite{des,fle}
\begin{eqnarray}
&&\ c_1^\prime=c_1, \quad  c_2^\prime=c_2,\nonumber\\
&&\ c_3^\prime=c_3-P_s/3, \quad  c_4^\prime=c_4+P_s,\nonumber\\
&&\ c_5^\prime=c_5-P_s/3, \quad  c_6^\prime=c_6+P_s,\nonumber\\
&&\ c_7^\prime=c_7+P_e, \quad  c_8^\prime=c_8,\nonumber\\
&&\ c_9^\prime=c_9+P_e, \quad  c_{10}^\prime=c_{10}, \label{c'c}
\end{eqnarray} where
\begin{eqnarray}
&&\ P_s=(\alpha_s/8\pi)c_2[10/9+G(m_c,\mu,q^2)],\nonumber\\
&&\
P_e=(\alpha_{em}/9\pi)(3c_1+c_2)[10/9+G(m_c,\mu,q^2)],\nonumber
\end{eqnarray}
with
\begin{eqnarray}
G(m_c,\mu,q^2)=4\int_{0}^{1}{\rm{d}}xx(x-1){\ln}\frac{m_c^2-x(1-x)q^2}{\mu^2},
\nonumber
\end{eqnarray}
where $m_c$ is the $c$-quark mass and $q^2$ is the typical
momentum transfer of the gluon or photon in the penguin diagrams.
$G(m_c,\mu,q^2)$ has the following explicit expression \cite{kra}:
\begin{eqnarray}
{\mathfrak{Re}}G&=&\frac{2}{3}\Bigg({\ln}\frac{m_c^2}{\mu^2}-\frac{5}{3}-4\frac{m_c^2}{q^2}+\bigg(1+2\frac{m_c^2}{q^2}\bigg)
\sqrt{1-4\frac{m_c^2}{q^2}}{\ln}\frac{1+\sqrt{1-4\frac{m_c^2}{q^2}}}{1-\sqrt{1-4\frac{m_c^2}{q^2}}}\Bigg),\nonumber\\
{\mathfrak{Im}}G&=&-\frac{2}{3}\pi\bigg(1+2\frac{m_c^2}{q^2}\bigg)\sqrt{1-4\frac{m_c^2}{q^2}}.\label{G}
\end{eqnarray}

The value of $q^2$ is chosen to be in the range
$0.3<q^2/m_b^2<0.5$ \cite{eno,gar}. From Eqs. (\ref{cvalue})
(\ref{c'c}) (\ref{G}) we can obtain numerical values of
$c_i^\prime$ which are listed in Table \ref{wilson coefficients},
where we have taken $\alpha_s(m_Z)$=0.118,
$\alpha_{em}(m_b)$=1/132.2, $m_b$=5 GeV, and $m_c$=1.35 GeV.
\begin{table*}[htb]
\caption{\label{table1}Effective Wilson coefficients for the tree
operators, electroweak and QCD penguin operators \cite{fle,kra}}
\begin{tabular}{c c c}
\hline
$c_i^\prime$ & $q^2/m_b^2$=0.3 & $q^2/m_b^2$=0.5\\
\hline
$c_1^\prime$ &\qquad $-0.3125$       &\quad $-0.3125$\\
$c_2^\prime$ &\qquad $1.1502$       &\quad $1.1502$\\
$c_3^\prime$ &\qquad $2.433 \times 10^{-2}+1.543\times 10^{-3}i$ &\quad  $2.120\times10^{-2}+5.174\times 10^{-3}i$\\
$c_4^\prime$ &\qquad $-5.808 \times 10^{-2}-4.628\times 10^{-3}i$ &\quad  $-4.869\times10^{-2}-1.552\times10^{-2}i$\\
$c_5^\prime$ &\qquad $1.733\times10^{-2}+1.543\times10^{-3}i$  &\quad  $1.420\times10^{-2}+5.174\times10^{-3}i$\\
$c_6^\prime$ &\qquad $-6.668\times10^{-2}-4.628\times10^{-3}i$  &\quad  $-5.729\times10^{-2}-1.552\times10^{-2}i$\\
$c_7^\prime$ &\qquad $-1.435\times10^{-4}-2.963\times10^{-5}i$  &\quad  $-8.340\times10^{-5}-9.938\times10^{-5}i$\\
$c_8^\prime$ &\qquad $3.839\times10^{-4}$  &\quad $3.839\times10^{-4}$\\
$c_9^\prime$ &\qquad $-1.023\times10^{-2}-2.963\times10^{-5}i$  &\quad  $-1.017\times10^{-2}-9.938\times10^{-5}i$\\
$c_{10}^\prime$  &\qquad $1.959\times10^{-3}$ &\quad  $1.959\times10^{-3}$\\
\hline\label{wilson coefficients}
\end{tabular}
\end{table*}

The CKM matrix, which should be determined from experiments, can
be expressed in terms of the Wolfenstein parameters,
$A,\lambda,\rho$ and $\eta$ \cite{wol}:

\begin{eqnarray}
\left(\begin{array}{ccc}1-\frac{1}{2}\lambda^2 & \lambda &
A\lambda^3(\rho-i\eta)\\-\lambda & 1-\frac{1}{2}\lambda^2 &
A\lambda^2\\ A\lambda^3(1-\rho-i\eta) & -A\lambda^2 & 1
\end{array}\right),
\end{eqnarray}
where $O(\lambda^4)$ corrections are neglected. The latest values
for the parameters in the CKM matrix are \cite{W.-M. Yao}:
\begin{eqnarray}
&& \lambda=0.2272\pm0.0010,\quad A=0.818_{-0.017}^{+0.007},\nonumber \\
&& \bar{\rho}=0.221_{-0.028}^{+0.064},\quad
\bar{\eta}=0.340_{-0.045}^{+0.017},\label{eq: rhobarvalue}
\end{eqnarray}
where
\begin{eqnarray}
 \bar{\rho}=\rho(1-\frac{\lambda^2}{2}),\quad
\bar{\eta}=\eta(1-\frac{\lambda^2}{2}).\label{eq: rho rhobar
relation}
\end{eqnarray}
From Eqs. (\ref{eq: rhobarvalue}) ( \ref{eq: rho rhobar relation})
we have
\begin{eqnarray}
0.198<\rho<0.293,\quad  0.302<\eta<0.366.\label{eq: rho value}
\end{eqnarray}
\section{\label{sec:cpv1}$CP$ violation in $\bar{B}^0\rightarrow\rho^0(\omega)\rho^0(\omega)\rightarrow \pi^+\pi^-\pi^+\pi^-$}
\subsection{\label{subsec:form}Formalism}

 Letting $A$ ($\bar{A}$) be the amplitude for the decay $\bar{B}^0\rightarrow
\pi^+\pi^-\pi^+\pi^-$ ($B^0\rightarrow \pi^+\pi^-\pi^+\pi^-$) one
has:
\begin{eqnarray}
A=\big<\pi^+\pi^-\pi^+\pi^-|H^T|\bar{B}^0\big>+\big<\pi^+\pi^-\pi^+\pi^-|H^P|\bar{B}^0\big>,\label{A}\\
\bar{A}=\big<\pi^+\pi^-\pi^+\pi^-|H^T|B^0\big>+\big<\pi^+\pi^-\pi^+\pi^-|H^P|B^0\big>,
\end{eqnarray}
with $H^T$ and $H^P$ being the Hamiltonian for the tree and
penguin operators, respectively.

We can define the relative magnitude and phases between the tree
and penguin operator contributions as follows:
\begin{eqnarray}
A=\big<\pi^+\pi^-\pi^+\pi^-|H^T|\bar{B}^0\big>[1+re^{i(\delta+\phi)}],\label{A'}\\
\bar{A}=\big<\pi^+\pi^-\pi^+\pi^-|H^T|B^0\big>[1+re^{i(\delta-\phi)}],
\label{Abar}
\end{eqnarray}
where $\delta$ and $\phi$ are strong and weak relative phases,
respectively. The phase $\phi$ can be expressed as a combination
of the CKM matrix elements:
$\phi={\rm{arg}}[(V_{tb}V_{td}^*)/(V_{ub}V_{ud}^*)]$. As a result,
sin$\phi$ is equal to sin$\alpha$ with $\alpha$ being defined in
the standard way \cite{W.-M. Yao}. The parameter $r$ is the
absolute value of the ratio of penguin and tree amplitudes:
\begin{eqnarray}
r\equiv\Bigg|\frac{\big<\pi^+\pi^-\pi^+\pi^-|H^P|\bar{B}^0\big>}{\big<\pi^+\pi^-\pi^+\pi^-|H^T|\bar{B}^0\big>}\Bigg|
\label{r}.
\end{eqnarray}
The $CP$ violating asymmetry, $a$, can be written as
\begin{eqnarray}
a\equiv\frac{|A|^2-|\bar{A}|^2}{|A|^2+|\bar{A}|^2}=\frac{-2r
{\rm{sin}}\delta {\rm{sin}}\phi}{1+2r {\rm{cos}}\delta
{\rm{cos}}\phi+r^2}. \label{asy}
\end{eqnarray}
In order to obtain a large signal for direct $CP$ violation, we
need some mechanism to make sin$\delta$ large. It has been found
that $\rho-\omega$ mixing has the dual advantages that it leads to
a large strong phase difference and is well known
\cite{gar,guo1,lei,guo2}. With this mechanism, to the first order
of isospin violation, we have the following results when the
invariant masses of $\pi^+\pi^-$ pairs are near the $\omega$
resonance mass:
\begin{eqnarray}
\big<\pi^+\pi^-\pi^+\pi^-|H^T|\bar{B}^0\big>=\frac{2g_{\rho}^2}{s_{\rho}^{2}s_{\omega}}\widetilde{\Pi}_{\rho\omega}t_{\rho\omega}+\frac{g_{\rho}^2}{s_{\rho}^2}t_{\rho\rho},
\label{Htr}\\
\big<\pi^+\pi^-\pi^+\pi^-|H^P|\bar{B}^0\big>=\frac{2g_{\rho}^2}{s_{\rho}^{2}s_{\omega}}\widetilde{\Pi}_{\rho\omega}p_{\rho\omega}+\frac{g_{\rho}^2}{s_{\rho}^2}p_{\rho\rho}.
\label{Hpe}
\end{eqnarray}
Here $t_{\rho\rho}(p_{\rho\rho})$ and
$t_{\rho\omega}(p_{\rho\omega})$ are the tree (penguin) amplitudes
for $\bar{B}\rightarrow\rho^0\rho^0$ and
$\bar{B}^0\rightarrow\rho^0\omega$, respectively, $g_{\rho}$ is
the coupling for $\rho^0\rightarrow\pi^+\pi^-$,
$\widetilde{\Pi}_{\rho\omega}$ is the effective $\rho-\omega$
mixing amplitude which also effectively includes the direct
coupling $\omega\rightarrow\pi^+\pi^-$, and $s_{V}$($V$=$\rho$ or
$\omega$) is the inverse propagator of the vector meson $V$,
\begin{eqnarray}
s_V=s-m_V^2+{\rm{i}}m_V\Gamma_V,
\end{eqnarray}
with $\sqrt{s}$ being the invariant masses of the $\pi^+\pi^-$
pairs (we let the invariant masses of the two $\pi^+\pi^-$ pairs
be the same). Eqs. (\ref{Htr}) (\ref{Hpe}) have different forms
from the case where only single $\rho-\omega$ mixing is involved
\cite{guo1,guo2,lei}: there is a factor of 2 in front of the
effective $\rho-\omega$ mixing amplitude,
$\widetilde{\Pi}_{\rho\omega}$, since $\rho-\omega$ mixing
contributes twice to the first order of isospin violation.
Furthermore, we have $g_\rho^2$ and $s_\rho^2$ instead of $g_\rho$
and $s_\rho$ as before due to two $\rho\rightarrow\pi\pi$
couplings and two $\rho$ propagators (note that $s_{\omega}^2$
term is of the second order of isospin violation and hence is
ignored).

As mentioned before, the direct coupling $\omega\rightarrow\pi^+\pi^-$ has been
effectively absorbed into $\widetilde{\Pi}_{\rho\omega}$
\cite{oco}. This leads to the explicit $s$ dependence of
$\widetilde{\Pi}_{\rho\omega}$. In practice, however, the $s$ dependence
of $\widetilde{\Pi}_{\rho\omega}$ is negligible. Making the expansion
$\widetilde{\Pi}_{\rho\omega}(s)=\widetilde{\Pi}_{\rho\omega}(m_{\omega}^2)+(s-m_{\omega})\widetilde{\Pi}_{\rho\omega}^\prime(m_{\omega}^2)$,
the $\rho-\omega$ mixing parameters were determined in the
fit of Gardner and O'Connell \cite{gard}:
\begin{eqnarray}
\mathfrak{Re}\widetilde{\Pi}_{\rho\omega}(m_{\omega}^2)&=&-3500\pm300
\rm{MeV}^2,\nonumber\\{\mathfrak{Im}}\widetilde{\Pi}_{\rho\omega}(m_{\omega}^2)&=&-300\pm300
\textrm{MeV}^2,\nonumber\\\widetilde{\Pi}_{\rho\omega}^\prime(m_{\omega}^2)&=&0.03\pm0.04.
\end{eqnarray}
From Eqs. (\ref{A})(\ref{A'})(\ref{Htr})(\ref{Hpe}) one has
\begin{eqnarray}
re^{i\delta}e^{i\phi}=\frac{2\widetilde{\Pi}_{\rho\omega}p_{\rho\omega}+s_{\omega}p_{\rho\rho}}{2\widetilde{\Pi}_{\rho\omega}t_{\rho\omega}+s_{\omega}t_{\rho\rho}},
\label{rdtdirive}
\end{eqnarray}
where the factor of 2 in front of $\widetilde{\Pi}_{\rho\omega}$
arises from the involvement of double $\rho-\omega$ mixing.
Defining
\begin{eqnarray}
\frac{p_{\rho\omega}}{t_{\rho\rho}}\equiv r^\prime
e^{i(\delta_q+\phi)},\quad\frac{t_{\rho\omega}}{t_{\rho\rho}}\equiv
\alpha
e^{i\delta_\alpha},\quad\frac{p_{\rho\rho}}{p_{\rho\omega}}\equiv
\beta e^{i\delta_\beta}, \label{def}
\end{eqnarray}
where $\delta_\alpha$, $\delta_\beta$ and $\delta_q$ are strong
phases, one finds the following expression from Eqs.
(\ref{rdtdirive})(\ref{def}):
\begin{eqnarray}
re^{i\delta}=r^\prime
e^{i\delta_q}\frac{2\widetilde{\Pi}_{\rho\omega}+\beta
e^{i\delta_\beta}s_{\omega}}{2\widetilde{\Pi}_{\rho\omega}\alpha
e^{i\delta_\alpha}+s_{\omega}}. \label{rdt}
\end{eqnarray}
In order to obtain the $CP$ violating asymmetry in Eq.
(\ref{asy}), sin$\phi$ and cos$\phi$ are needed, where $\phi$ is
determined by the CKM matrix elements. In the Wolfenstein
parametrization \cite{wol}, one has
\begin{eqnarray}
\sin\phi=\frac{\eta}{\sqrt{[\rho(1-\rho)-\eta^2]^2+\eta^2}},\\
\cos\phi=\frac{\rho(1-\rho)-\eta^2}{\sqrt{[\rho(1-\rho)-\eta^2]^2+\eta^2}}.
\end{eqnarray}

\subsection{\label{subsec:cal}Calculational details}

With the Hamiltonian given in Eq. (\ref{ham}) we can evaluate the
matrix elements for
$\bar{B}^0\rightarrow\rho^0(\omega)\rho^0(\omega)$. In the
factorization approximation, $\rho^0(\omega)$ is generated by one
current which has the appropriate quantum numbers in the
Hamiltonian. For this decay process, the amplitude can be written
as the product of two matrix elements after factorization, i.e.
(omitting Dirac matrices and color labels):
$\langle\rho^0(\omega)|(\bar{q}q)|0\rangle\langle\rho^0(\omega)|(\bar{d}b)|\bar{B}^0\rangle$
($q=u,d$), where $(\bar{q}q)$ and $(\bar{d}b)$ denote the $V-A$
currents, $\bar{q}\gamma_{\mu}(1-\gamma_5)q$ and
$\bar{d}\gamma_{\mu}(1-\gamma_5)b$, respectively. Since $\rho^0$
and $\omega$ are vector mesons the amplitude for
$\bar{B}^0\rightarrow\rho^0(\omega)\rho^0(\omega)$ may be
polarized or unpolarized. Here we investigate the later case.
Defining
\begin{eqnarray}
\langle\rho^0|(\bar{u}u)|0\rangle\langle\rho^0|(\bar{d}b)|\bar{B}^0\rangle\equiv
T,
\end{eqnarray}
one has
\begin{eqnarray}
T&&=-\langle\rho^0|(\bar{d}d)|0\rangle\langle\rho^0|(\bar{d}b)|\bar{B}^0\rangle\nonumber\\&&=-\langle\rho^0|(\bar{u}u)|0\rangle\langle\omega|(\bar{d}b)|\bar{B}^0\rangle\nonumber\\
&&=\langle\rho^0|(\bar{d}d)|0\rangle\langle\omega|(\bar{d}b)|\bar{B}^0\rangle\nonumber\\&&=\langle\omega|(\bar{u}u)|0\rangle\langle\rho^0|(\bar{d}b)|\bar{B}^0\rangle.
\end{eqnarray}
After factorization, the contribution to $t_{\rho\rho}$ from the
tree level operator $O_1^u$ is
\begin{eqnarray}
\langle\rho^0\rho^0|O_1^u|\bar{B}^0\rangle=2\langle\rho^0|(\bar{u}u)|0\rangle\langle\rho^0|(\bar{d}b)|\bar{B}^0\rangle=2T.
\end{eqnarray}
Using the Fierz transformation the contribution of $O_2^u$ is
$(1/N_c)T$. Hence we have
\begin{eqnarray}
t_{\rho\rho}=2\bigg(c_1^\prime+\frac{1}{N_c}c_2^\prime\bigg)T.
\label{tro}
\end{eqnarray}

It should be noted that in Eq. (\ref{tro}) we have neglected the
color-octet contribution which is nonfactorizable and difficult to
calculate. Therefore, $N_c$ should be treated as an effective
parameter and may deviate from the naive value 3
\cite{guo1,guo2,lei}. In the same way we find that
$t_{\rho\omega}=0$. This lead to
\begin{equation}
\alpha e^{i\delta_\alpha}=0,
\end{equation}
from Eq. (\ref{def}).

In a similar way, we can evaluate the penguin operator
contributions $p_{\rho\rho}$ and $p_{\rho\omega}$ with the aid of
the Fierz identities. From Eq. (\ref{def}) we have

\begin{eqnarray} \beta
e^{i\delta_\beta}\!\!&=\!\!&\frac{-2\Big(c_4^\prime+\frac{1}{N_c}c_3^\prime\Big)+3\Big(c_7^\prime+\frac{1}{N_c}c_8^\prime\Big)+3\Big(c_9^\prime+\frac{1}{N_c}c_{10}^\prime\Big)+\Big(c_{10}^\prime+\frac{1}{N_c}c_9^\prime\Big)}{2\Big(c_3^\prime+\frac{1}{N_c}c_4^\prime\Big)+2\Big(c_4^\prime+\frac{1}{N_c}c_3^\prime\Big)+2\Big(c_5^\prime+\frac{1}{N_c}c_6^\prime\Big)-\Big(c_7^\prime+\frac{1}{N_c}c_8^\prime\Big)-\Big(c_{9}^\prime+\frac{1}{N_c}c_{10}^\prime\Big)-\Big(c_{10}^\prime+\frac{1}{N_c}c_9^\prime\Big)},\nonumber\\
\\ r^\prime
e^{i\delta_q}\!\!\!&=\!\!&\frac{-2\Big(c_3^\prime+\frac{1}{N_c}c_4^\prime\Big)-2\Big(c_4^\prime+\frac{1}{N_c}c_3^\prime\Big)-2\Big(c_5^\prime+\frac{1}{N_c}c_6^\prime\Big)+\Big(c_7^\prime+\frac{1}{N_c}c_8^\prime\Big)+\Big(c_{9}^\prime+\frac{1}{N_c}c_{10}^\prime\Big)+\Big(c_{10}^\prime+\frac{1}{N_c}c_9^\prime\Big)}{2\bigg(c_1^\prime+\frac{1}{N_c}c_2^\prime\bigg)}\nonumber\\&
& \times\bigg|\frac{V_{tb}V_{td}^*}{V_{ub}V_{ud}^*}\bigg|,
\end{eqnarray}
where
\begin{eqnarray}
\bigg|\frac{V_{tb}V_{td}^*}{V_{ub}V_{ud}^*}\bigg|=\frac{\sqrt{(1-\rho)^2+\eta^2}}{(1-\lambda^2/2)\sqrt{\rho^2+\eta^2}}.
\end{eqnarray}

\subsection{\label{sec:numdis}Numerical results}

We have several parameters in the numerical calculations: $q^2$,
$N_c$, and the CKM matrix elements. Since $N_c$ includes
nonfactorizable effects, which cannot be evaluated accurately at
present, we choose to treat it as a parameter and determine its
range from the experimental data. Then, we can extract an allowed
range for $N_c$ from a comparison of the theoretical results and
the experimental data. By doing this, we get the range of $N_c$ as
$2.74 (2.81)<N_c<4.77 (4.92)$ for $q^2/m_b^2=0.3 (0.5)$. This will
be discussed in detail in Sec. \ref{sec:branchratio}. The most
uncertainties due to the CKM matrix elements come from $\rho$ and
$\eta$ since $\lambda$ is well determined (see Eq. (\ref{eq:
rhobarvalue})) and since the $CP$ violating asymmetry is
independent of the Wolfenstein parameter $A$. Therefore, in our
numerical calculations, we take the central value for $\lambda$
and only let $(\rho, \eta)$ vary between the limiting values
$(\rho_{min}, \eta_{min})$ and $(\rho_{max}, \eta_{max})$. In
fact, explicit numerical results show that the $CP$ violating
asymmetry is very insensitive to $\lambda$.

In the numerical calculations, it is found that for a fixed $N_c$
there is a maximum value, $a_{max}$, for the $CP$ violating
parameter, $a$, when the invariant masses of the $\pi^+\pi^-$
pairs are in the vicinity of the $\omega$ resonance. This is shown
explicitly in Fig. 1. For $q^2/m_b^2=0.3(0.5)$ and
$N_c=2.74(2.81)$, the maximum $CP$ violating asymmetry varies from
around -91.1\% (-70.1\%) to around -96.1\% (-77.8\%) as
$(\rho,\eta)$ change from $(\rho _{max}, \eta_{max})$ to
$(\rho_{min}, \eta_{min})$;  For $q^2/m_b^2=0.3(0.5)$ and
$N_c=4.77(4.92)$, the maximum $CP$ violating asymmetry varies from
around 55.8\% (28.9\%) to around 53.2\% (22.9\%) when
$(\rho,\eta)$ change from $(\rho _{max}, \eta_{max})$ to
$(\rho_{min}, \eta_{min})$.

Our results show that the $\rho-\omega$ mixing mechanism produces
a large sin$\delta$ in the allowed range of $N_c$, which is
necessary for a large $CP$ violating asymmetry. The involvement of
double $\rho-\omega$ mixing in
$\bar{B}^0\rightarrow\rho^0(\omega)\rho^0(\omega)\rightarrow\pi^+\pi^-\pi^+\pi^-$
gives rise to a factor of 2 in Eq. (\ref{rdtdirive}) in front of
$\widetilde{\Pi}_{\rho\omega}$. This makes the $CP$ asymmetry even
larger than in the case where the $\rho-\omega$ mixing contributes
only once. Fig. 2 shows explicitly the comparison between these
two cases for $N_c=2.74 (2.81)$ and $q^2/m_b^2=0.3 (0.5)$. The
maximum asymmetry with the involvement of single $\rho-\omega$
mixing, for $q^2/m_b^2=0.3 (0.5)$ and $N_c=2.74 (2.81)$, is around
-55.2\% (-20.0\%) for the set ($\rho_{max}, \eta_{max}$) and
-63.2\% (-23.8\%) for the set ($\rho_{min}\eta_{min}$). For
$q^2/m_b^2=0.3 (0.5)$ and $N_c=4.77 (4.92)$, we find that
$a_{max}$ is around 26.5\% (-1.49\%) for the set ($\rho_{max},
\eta_{max}$) and 25.1\% (-1.50\%) for the set
($\rho_{min}\eta_{min}$) in the case of single $\rho-\omega$
mixing.

The reason that double $\rho-\omega$ mixing leads to a larger $CP$
violating asymmetry than in the case of single $\rho-\omega$
mixing is that sin$\delta$ becomes bigger in the case of double
$\rho-\omega$ mixing than in the case of single $\rho-\omega$
mixing. This can be seen explicitly from Fig. 3. The involvement
of double $\rho-\omega$ mixing may also change the value of $r$ as
can be seen from Fig. 4. However, as found from our detailed
analysis for the influence of $r$ on the $CP$ violating asymmetry,
the effect of the change of $r$ on $a$ is small compared with the
change of sin$\delta$ due to the involvement of double
$\rho-\omega$ mixing.

It is noted that when $N_c$ is around 2.81 and 4.92 in the case
$q^2/m_b^2=0.5$, we could also have large $CP$ asymmetries when
$\sqrt{s}$ is far away from the $\omega$ resonance for all the
allowed values of the CKM matrix elements (see Figs. 1(b) and
5(b)). In these cases, the effective $\rho-\omega$ mixing
contributes little and the large $CP$ asymmetry is caused by the
effective Wilson coefficients, which can also give a large strong
phase, $\delta$, since they are complex numbers.

In most direct $CP$ violating decays such as
$\bar{B}^0\rightarrow\rho^0(\omega)\pi^0\rightarrow\pi^+\pi^-\pi^0$
\cite{guo2,lei} and some other processes, the involvement of
$\rho-\omega$ mixing leads to the result that the strong phase,
$\delta$, passes through $90^{\circ}$ (sin$\delta$=1) at the
$\omega$ resonance. However, in the decay we are discussing this
does not happen in the allowed range of $N_c$. Instead, the
absolute value of sin$\delta$ just gets close to 1, but does not
equal 1 (see Fig. 3), even though it is enough to give large $CP$
asymmetry, especially when double $\rho-\omega$ mixing is
involved.

Figs. 5 and 6 show the dependence of the $CP$ violating asymmetry
and sin$\delta$, respectively, on both $N_c$ and $\sqrt s$. One
can see that the $CP$ asymmetry strongly depends on $N_c$. Take
Fig. 5(a) as an example (for $q^2/m_b^2=0.3$ and maximum $\rho$
and $\eta$): when $N_c<3.68$, one gets minus asymmetry around the
$\omega$ resonance, whereas when $N_c>3.68$ the $CP$ violating
asymmetry becomes positive.

It can be seen from Fig. 5 that when $N_c$ takes the critical
value, $-c_2^{\prime}/c_1^{\prime}\simeq3.68$, the $CP$ violating
asymmetry becomes zero. This is because $t_{\rho\rho}=0$ at this
point (as can be seen from Eq. (\ref{tro}) easily) and hence the
penguin operator contributions dominate. Furthermore, the sign of
sin$\delta$ and hence the sign of the $CP$ violating asymmetry
change at this point. It would be interesting to see whether or
not $N_c$ can take this value in the future when more accurate
experimental date are available. From most previous studies, it
seems that $N_c$ is usually less than this critical value
\cite{gar,guo1,guo2,lei,chen}. If this is true for $\bar{B}^0
\rightarrow\rho^0(\omega)\rho^0(\omega)$, then the sign of
sin$\delta$ would remain unchanged. Then, one could remove the mod
($\pi$) ambiguity in the determination of the $CP$ violating phase
angle $\alpha$ (through sin2$\alpha$) by measuring the $CP$
violating asymmetry in $\bar{B}^0
\rightarrow\rho^0(\omega)\rho^0(\omega)\rightarrow\pi^+\pi^-\pi^+\pi^-$.

\section{\label{sec:branchratio}Branching ratios for $\bar{B}^0 \rightarrow\rho^0\rho^0$ and
$\bar{B}^0 \rightarrow\rho^+\rho^-$}

\subsection{\label{sec:brformalism}Formalism}
If the decay amplitude for $B\rightarrow V_1V_2$ ($V_1, V_2$
denote vector mesons) has the form $A(B\rightarrow V_1V_2)=\alpha
X^{(BV_1,V_2)}$, where $X^{(BV_1,V_2)}$ denotes the factorizable
amplitude with the form $\langle
V_2|(\bar{q}_2q_3)|0\rangle\langle V_1|(\bar{q}_1b)|B\rangle$,
then the decay rate is given by \cite{chen}
\begin{equation}
\Gamma(B\rightarrow V_1V_2)=\frac{p_c}{8\pi m_B^2}|\alpha
(m_B+m_1)m_2f_{V_2}A_1^{BV_1}(m_2)|^2H,\label{eq:gamma BtoV1V2}
\end{equation}
where $\alpha$ is related to the CKM matrix elements and Wilson
coefficients, $f_{V_2}$ is the decay constant of $V_2$, $p_c$ is
the c.m. momentum of the decay particles, $m_B$ and $m_1(m_2)$ are
the masses of the $B$ meson and the vector meson $V_1(V_2)$,
respectively, and
\begin{equation}
H=(a-bx)^2+2(1-c^2y^2)\label{eq:parameter H},
\end{equation}
where
\begin{eqnarray}
&&a=\frac{m_B^2-m_1^2-m_2^2}{2m_1m_2},\;\;b=\frac{2m_B^2p_c^2}{m_1m_2(m_B+m_1)^2},\;\;c=\frac{2m_Bp_c}{(m_B+m_1)^2},\nonumber\\
&&x=\frac{A_2^{BV_1}(m_2^2)}{A_1^{BV_1}(m_2^2)},\;\;y=\frac{V^{BV_1}(m_2^2)}{A_1^{BV_1}(m_2^2)},\nonumber\\
&&p_c=\frac{\sqrt{[m_B^2-(m_1+m_2)^2][m_B^2-(m_1-m_2)^2]}}{2m_B}.\label{eq:
parameters in H}\label{eq: c.m. momentum}
\end{eqnarray}
$A_1^{BV_1}$, $A_2^{BV_1}$ and $V^{BV_1}$ in Eqs.
(\ref{eq:parameter H}) and (\ref{eq: parameters in H}) are the
form factors associated with $B\rightarrow V_1$ transition.

The decay amplitudes for $\bar{B}^0\rightarrow\rho^0\rho^0$,
$\bar{B}^0\rightarrow\rho^0\omega$ and $\bar{B}^0\rightarrow
\rho^+\rho^-$ are
\begin{equation}
A(\bar{B}^0\rightarrow\rho^0\rho^0)=\alpha_1X^{(B\rho^0,\rho^0)},\label{eq:
amplitude B0 to rho0 rho0}
\end{equation}
\begin{equation}
A(\bar{B}^0\rightarrow\rho^0\omega)=\alpha_2X^{(B\rho^0,\omega)},\label{eq:
amplitude B0 to rho0 omega}
\end{equation}
and
\begin{equation}
A(\bar{B}^0\rightarrow\rho^+\rho^-)=\alpha_3X^{(B\rho^+,\rho^-)},\label{eq:
amplitude B0 to rho+ rho-}
\end{equation}
where
\begin{equation}
\alpha_1=\frac{G_F}{\sqrt{2}}[2a_1V_{ub}V_{ud}^*-(-2a_4+3a_7+3a_9+a_{10})V_{tb}V_{td}^*],\label{eq:
alpha 1}
\end{equation}
\begin{equation}
\alpha_2=-\frac{G_F}{\sqrt{2}}(2a_3+2a_4+2a_5-a_7-a_9-a_{10})V_{tb}V_{td}^*,\label{eq:
alpha 2}
\end{equation}
\begin{equation}
\alpha_3=-\frac{G_F}{\sqrt{2}}[a_2V_{ub}V_{ud}^*-(a_4+a_{10})V_{tb}V_{td}^*],\label{eq:
alpha 3}
\end{equation}
with $a_i$ ($i= 1, 2, \cdots,10$) being defined as:
\begin{eqnarray}
&&a_{2j}=c_{2j}^\prime+\frac{c_{2j-1}^\prime}{N_c},\nonumber\\
&&a_{2j-1}=c_{2j-1}^\prime+\frac{c_{2j}^\prime}{N_c},\;\;\textrm{for}
\;\;j=1,2,\cdots,5.\label{eq: parameters a_i}
\end{eqnarray}

When we calculate the branching ratios we should take into account
the $\rho-\omega$ mixing contribution for consistency since we are
working to the first order of isospin violation. Then, we obtain
the branching ratio for $\bar{B}^0 \rightarrow\rho^0\rho^0$:
\begin{equation}
BR(\bar{B}^0 \rightarrow\rho^0\rho^0)=\frac{p_c}{8\pi
m_B^2\Gamma_{B^0}}\left|\left(\alpha_1+\alpha_2\frac{2\widetilde{\Pi}_{\rho\omega}}{(s_{\rho}-m_{\omega}^2)+\textrm{i}
m_{\omega}\Gamma_{\omega}}\right)
(m_B+m_{\rho^0})m_{\rho^0}f_{\rho^0}A_1(m_{\rho^0}^2)\right|^2H.\label{eq:
branching ratio for B to rho0rho0}
\end{equation}
For $\bar{B}^0\rightarrow \rho^+\rho^-$, we have
\begin{equation}
BR(\bar{B}^0 \rightarrow\rho^+\rho^-)=\frac{p_c}{8\pi
m_B^2\Gamma_{B^0}}|(\alpha_3
(m_B+m_{\rho^+})m_{\rho^-}f_{\rho^-}A_1(m_{\rho^+}^2)|^2H.\label{eq:
branching ratio for B to rho+rho-}
\end{equation}

\subsection{\label{sec:brformfactormodels}Form factor models}
The form factors $A_1(k^2)$, $A_2(k^2)$ and $V(k^2)$ depend on the
inner structure of hadrons and consequently depend on the
phenomenological models for hadronic wave functions. We adopt the
following form factor models:

Model 1(2) \cite{bauer,guo3}:\\
\begin{equation}
V(k^2)=\frac{V(0)}{1-k^2/(m_{1^-}^2)},
A_1(k^2)=\frac{A_1(0)}{1-k^2/(m_{1^+}^2)},
A_2(k^2)=\frac{A_2(0)}{1-k^2/(m_{1^+}^2)},\label{eq: model 1(2)}
\end{equation}
where $V(0)=0.33(0.395)$, $A_1(0)=A_2(0)=0.28(0.345)$,
$m_{1^-}=5.32$GeV, and $m_{1^+}=5.71$GeV.

Model 3(4) \cite{bauer,guo3,chen}:
\begin{equation}
V(k^2)=\frac{V(0)}{[1-k^2/(m_{1^-}^2)]^2},
A_1(k^2)=\frac{A_1(0)}{1-k^2/(m_{1^+}^2)},
A_2(k^2)=\frac{A_2(0)}{[1-k^2/(m_{1^+}^2)]^2},\label{eq: model
3(4)}
\end{equation}
where the form factors have double pole dependence and the
parameters take the same values as in Models 1 and 2.

Model 5 \cite{melikhov}:\\
for $V(k^2)$:
\begin{equation}
V(k^2)=\frac{V(0)}{(1-k^2/m_V)[1-\sigma_1k^2/m_V^2+\sigma_2k^4/m_V^4]},\label{eq:
model 5-1}
\end{equation}
for $A_i(k^2)$ ($i$=1, 2):
\begin{equation}
A_i(k^2)=\frac{A_i(0)}{1-\sigma_1k^2/m_V^2+\sigma_2k^4/m_V^4},\label{eq:
model 5-2}
\end{equation}
where $m_V=m_{B^*}=5.32\textrm{GeV}$; $V(0)=0.31$, $\sigma_1=0.59$
and $\sigma_2=0$ for $V(k^2)$; $A_1(0)=0.26$, $\sigma_1=0.73$ and
$\sigma_2=0.10$ for $A_1(k^2)$; and $A_2(0)=0.24$, $\sigma_1=1.40$
and $\sigma_2=0.50$ for $A_2(k^2)$.

Model 6 \cite{ball1,ball2}:\\
the form factors $A_1(k^2)$, $A_2(k^2)$ and$V(k^2)$ have the same
form:
\begin{equation}
f(k^2)=\frac{f(0)}{1-a_Fk^2/m_B^2+b_Fk^4/m_B^4},\label{eq: model
6}
\end{equation}
where $f$ could be $A_1$, $A_2$, or $V$. The parameters $f(0)$,
$a_F$ and $b_F$ for various form factors are:
 for $A_1$, $A_1(0)=0.261$, $a_F=0.29$, $b_F=-0.415$; for
$A_2$, $A_2(0)=0.223$, $a_F=0.93$, $b_F=-0.092$; and for $V$,
$V(0)=0.338$, $a_F=1.37$, $b_F=0.315$.

\subsection{\label{sec:brnumericalresult}Numerical results}
As mentioned before, $N_c$ includes the nonfactorizable effects
effectively, which cannot be handled well at present. Therefore,
we treat $N_c$ as a parameter to be determined by experimental
data. Usually $N_c$ is assumed to be universal for all decay
channels in the factorization approach. However, it certainly
could be different for different channels. Therefore, we choose to
determine the range of $N_c$ for
$\bar{B}^0\rightarrow\rho^0(\omega)\rho^0(\omega)$ from the
experimental data for the branching ratios for the decays
$\bar{B}^0\rightarrow\rho^0\rho^0$ and
$\bar{B}^0\rightarrow\rho^+\rho^-$ (we expect the nonfactorizable
contributions  and hence the values of $N_c$ in these two channels
are the same if isospin violation is ignored). In order to find
the range allowed for $N_c$ we use the latest experimental data
for the branching ratios for the two decay channels,
$\bar{B}^0\rightarrow\rho^0\rho^0$ and
$\bar{B}^0\rightarrow\rho^+\rho^-$ \cite{W.-M. Yao}:
\begin{eqnarray}
{\rm BR}(\bar{B}^0\rightarrow\rho^0\rho^0)<1.1\times10^{-6},\nonumber\\
{\rm
BR}(\bar{B}^0\rightarrow\rho^+\rho^-)=(2.5\pm0.4)\times10^{-5}.
\end{eqnarray}

We calculate the branching ratios for
$\bar{B}^0\rightarrow\rho^0\rho^0$ and
$\bar{B}^0\rightarrow\rho^+\rho^-$ with the formulae given in Eqs.
(\ref{eq:gamma BtoV1V2}) -- (\ref{eq: branching ratio for B to
rho+rho-}) in all the models for the weak form factors associated
with $\bar{B}^0\rightarrow\rho^0$ and
$\bar{B}^0\rightarrow\rho^+(\rho^-)$, which are mentioned in the
previous subsection. In addition to the dependence on $N_c$, these
two branching ratios also depend on the CKM matrix elements which
are parameterized by $\lambda$, $A$, $\rho$ and $\eta$, with the
experimental values of them being given in Eqs. (\ref{eq:
rhobarvalue}) (\ref{eq: rho value}). Since each of these
parameters has some uncertainty, we let each of them vary in its
allowed range when we calculate the branching ratios. Then, for
each set of the values for the parameters $\lambda$, $A$, $\rho$
and $\eta$, we obtain a range of $N_c$ which is allowed by the
experimental data for the branching ratios for both
$\bar{B}^0\rightarrow\rho^0\rho^0$ and
$\bar{B}^0\rightarrow\rho^+\rho^-$. This is shown in Figs. 7 and 8
for a special set of CKM matrix parameters when $q^2/m_b^2=0.3$.
Repeating this process for various sets of the values for
$\lambda$, $A$, $\rho$ and $\eta$ and taking the union of the
ranges of $N_c$ for all these sets, we find a range for $N_c$
which covers the whole range for these CKM matrix parameters. We
repeat this process for all the form factor models mentioned in
Eqs. (\ref{eq: model 1(2)}) -- (\ref{eq: model 6}) and obtain the
range of $N_c$ for each model as shown in Table \ref{tab: N_c
range}. Taking the union of all the ranges for these models we
finally find the maximum possible range for $N_c$: $2.74<N_c<4.77$
and $2.81<N_c<4.92$ for $q^2/m_b^2=0.3$ and $q^2/m_b^2=0.5$,
respectively.

\begin{table*}[htb]
\caption{\label{tab: N_c range}The range of $N_c$ for all the
models and the maximum range of $N_c$.}
\begin{tabular}{c c c}
\hline
 & $q^2/m_b^2$=0.3 & $q^2/m_b^2$=0.5\\
\hline
Model 1 &\qquad $(2.74, 4.74)$       &\quad  $(2.81, 4.90)$\\
Model 2 &\qquad $(2.78, 4.47)$       &\quad  $(2.84, 4.64)$\\
Model 3 &\qquad $(2.75, 4.77)$       &\quad  $(2.81, 4.92)$\\
Model 4 &\qquad $(2.76, 4.54)$       &\quad  $(2.82, 4.72)$\\
Model 5 &\qquad $(2.74, 4.69)$       &\quad  $(2.81, 4.84)$\\
Model 6 &\qquad $(2.77, 4.50)$       &\quad  $(2.84, 4.68)$\\
maximum range &\qquad $(2.74, 4.77)$ &\quad  $(2.81, 4.92)$\\
\hline
\end{tabular}
\end{table*}

\section{\label{sec:conclusion}Summary and Discussion}

We have calculated the $CP$ violating asymmetry in the process
$\bar{B}^0\rightarrow\rho^0(\omega)\rho^0(\omega)\rightarrow\pi^+\pi^-\pi^+\pi^-$
including the effect of $\rho-\omega$ mixing. The advantage of
$\rho-\omega$ mixing is that it makes the strong phase difference,
$\delta$, between the hadronic matrix elements of the tree and
penguin operators very large  at the $\omega$ resonance for a
fixed $N_c$. We have found that sin$\delta$ becomes large and
reaches the maximum point at the $\omega$ resonance. Consequently,
the $CP$ violating asymmetry reaches the maximum value  when the
invariant masses of the $\pi^+\pi^-$ pairs in the decay product
are in the vicinity of the $\omega$ resonance. Furthermore, since
there are two $\rho (\omega)$ mesons in the intermediate state,
$\rho-\omega$ mixing contributes twice when we work to the first
order of isospin violation. This leads to an even larger $CP$
violating asymmetry than in the case where only single
$\rho-\omega$ mixing is involved. This is unique for the process
$\bar{B}^0\rightarrow\rho^0(\omega)\rho^0(\omega)\rightarrow\pi^+\pi^-\pi^+\pi^-$.
As a result, the largest $CP$ violating asymmetry could be more
than 90\% for some values of $N_c$. This could be observed in the
future experiments at LHC. Now, we roughly estimate the
possibility to observe this $CP$ violating asymmetry. If the
branching ratio for $\bar{B}^0\rightarrow\rho^0\rho^0$ is of order
$10^{-6}$, then the number of $B^0\bar{B}^0$ pairs needed for
observing the $CP$ violating asymmetry (90\%) is roughly
$\frac{1}{BR(\bar{B}^0\rightarrow\rho^0\rho^0)}\frac{1}{a^2}\sim
10^6$ for $1\sigma$ signature and $10^7$ for $3\sigma$ signature
\cite{du}. It has been pointed out that at LHC, the number of
$B^0\bar{B}^0$ pairs could be around $4\times10^7$ (for ATLAS and
CMS) and $4\times10^5$ (for LHCb) per year \cite{schopper}.
Therefore, it is possible to observe the $CP$ violation for
$\bar{B}^0\rightarrow\rho^0(\omega)\rho^0(\omega)\rightarrow\pi^+\pi^-\pi^+\pi^-$
via the double $\rho-\omega$ mixing mechanism at LHC.

In the calculation, we need the Wilson coefficients for the tree
and penguin operators at the scale $m_b$. We work with the
renormalization scheme independent Wilson coefficients. We have
found that apart from the $\rho-\omega$ mixing mechanism, the
Wilson coefficients themselves could also give observable $CP$
violating asymmetry in some cases. The errors in the CKM matrix
elements lead to some uncertainties in the $CP$ violating
asymmetry. Even bigger uncertainties come from the hadronic matrix
elements of the tree and penguin operators due to the
nonperturbative QCD effects. We have worked in the factorization
approach, with the effective parameter $N_c$ being introduced to
account for the nonfactorizable effects. We have shown that the
$CP$ violating asymmetry in this decay process strongly depends on
the parameter $N_c$.

In order to determine the range of $N_c$ we have compared the
theoretical values and the experimental data for the branching
ratios for $\bar{B}^0 \rightarrow\rho^0\rho^0$ and $\bar{B}^0
\rightarrow\rho^+\rho^-$. We have found that the latest
experimental data constrain $N_c$ to be in the range $(2.74,
4.77)$ for $q^2/m_b^2$=0.3 and $(2.81, 4.92)$ for $q^2/m_b^2$=0.5,
respectively, when we let the CKM matrix elements vary in the
ranges determined by the current experiments. We have studied the
sign of ${\rm sin} \delta$ in the range of $N_c$ and found that
${\rm sin} \delta$ changes its sign at the point $N_c=3.68$. This
also leads to the change of the sign of the $CP$ violating
asymmetry. Due to the large errors in the current experimental
data for the branching ratios for $\bar{B}^0
\rightarrow\rho^0\rho^0$ and $\bar{B}^0 \rightarrow\rho^+\rho^-$
we cannot constrain $N_c$ more accurately at present. If the
future experimental data could constrain $N_c$ to be less than
3.68 ($N_c$ is usually less than 3.68 in other studies), the sign
of the $CP$ violating asymmetry would remain unchanged in the
whole range of $N_c$. Then one could remove the mod $(\pi)$
ambiguity in the determination of the $CP$ violating phase angle
$\alpha$ by measuring the $CP$ violating asymmetry in
$\bar{B}^0\rightarrow\rho^0(\omega)\rho^0(\omega)\rightarrow\pi^+\pi^-\pi^+\pi^-$.

For the decay process
$\bar{B}^0\rightarrow\rho^0(\omega)\rho^0(\omega)$, the
factorization approach we have used is expected to be a good
approximation since $B$ meson decays are energetic and since
$\alpha_s(m_b)$ and $1/m_b$ corrections should be small in the QCD
factorization scheme. One may also work in the QCD factorization
scheme, taking the value of $N_c$ to be 3 and including
corrections of order $\alpha_s(m_b)$ as done in Ref. \cite{lei2}.
However, the QCD factorization scheme suffers from endpoint
singularities which are not well controlled. The $CP$ violating
asymmetry depends on the unknown parameters which are associated
with such endpoint singularities. This lead to very uncertain $CP$
violating asymmetries in the QCD factorization scheme \cite{lei2}.
As mentioned before, the uncertainty for the $CP$ violating
asymmetry is also very large in the factorization approach we have
used, i.e. from about -96\% to about 56\% depending on the value
of $N_c$ and the CKM matrix elements. Furthermore, the $CP$
violating asymmetry may strongly depend on the factorization
approach adopted \cite{lei2}. All these issues need further and
more careful investigations.

\begin{acknowledgments}
This work was supported in part by National Natural Science
Foundation of China (Project Number 10675022), the Key Project of
Chinese Ministry of Education (Project Number 106024) and the
Special Grants from Beijing Normal University.
\end{acknowledgments}

\begin{figure}
\includegraphics[width=0.48\textwidth]{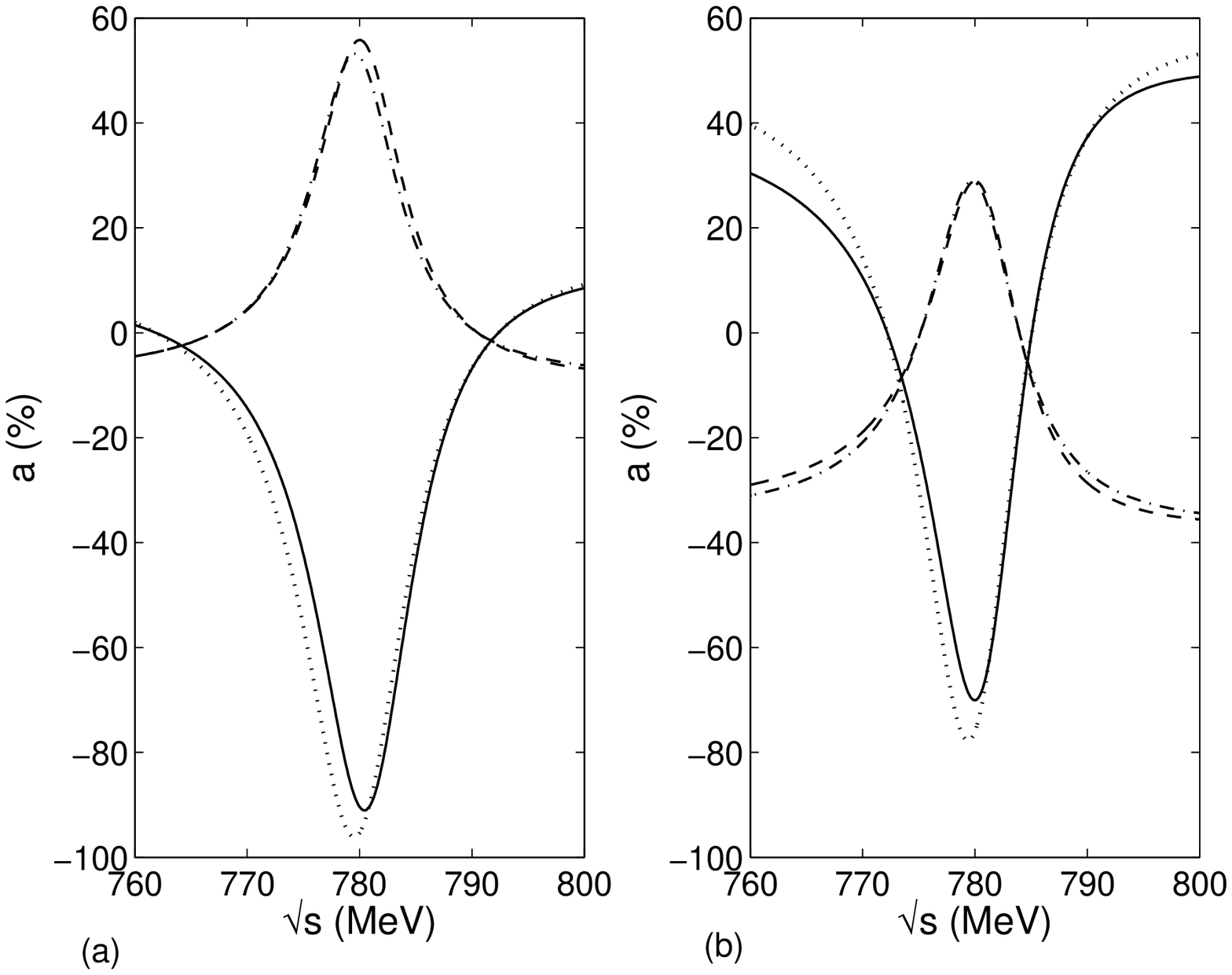}\\
\centerline{Fig. 1}
\end{figure}

\begin{figure}
\includegraphics[width=0.48\textwidth]{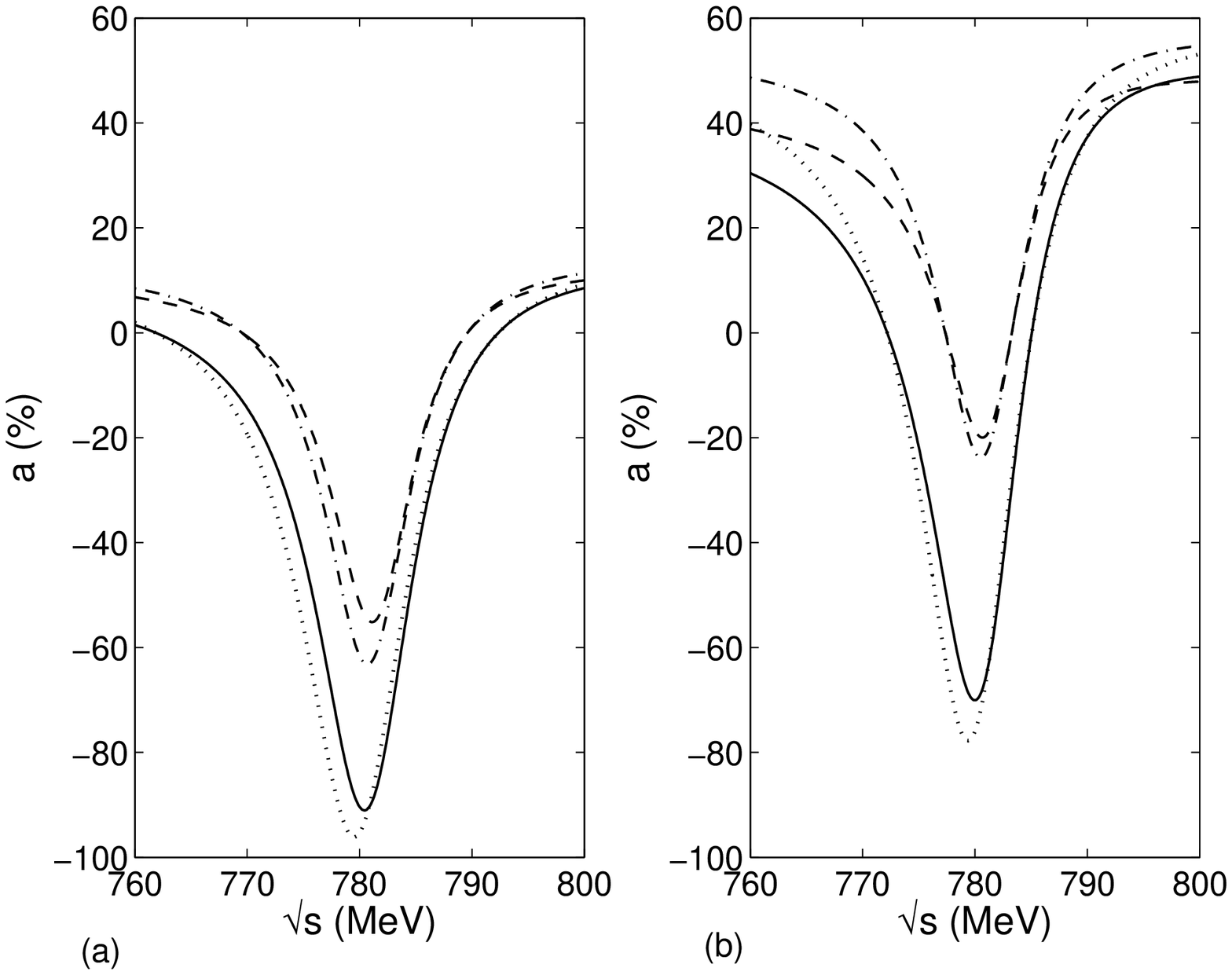}\\
\centerline{Fig. 2}
\end{figure}

\begin{figure}
\includegraphics[width=0.48\textwidth]{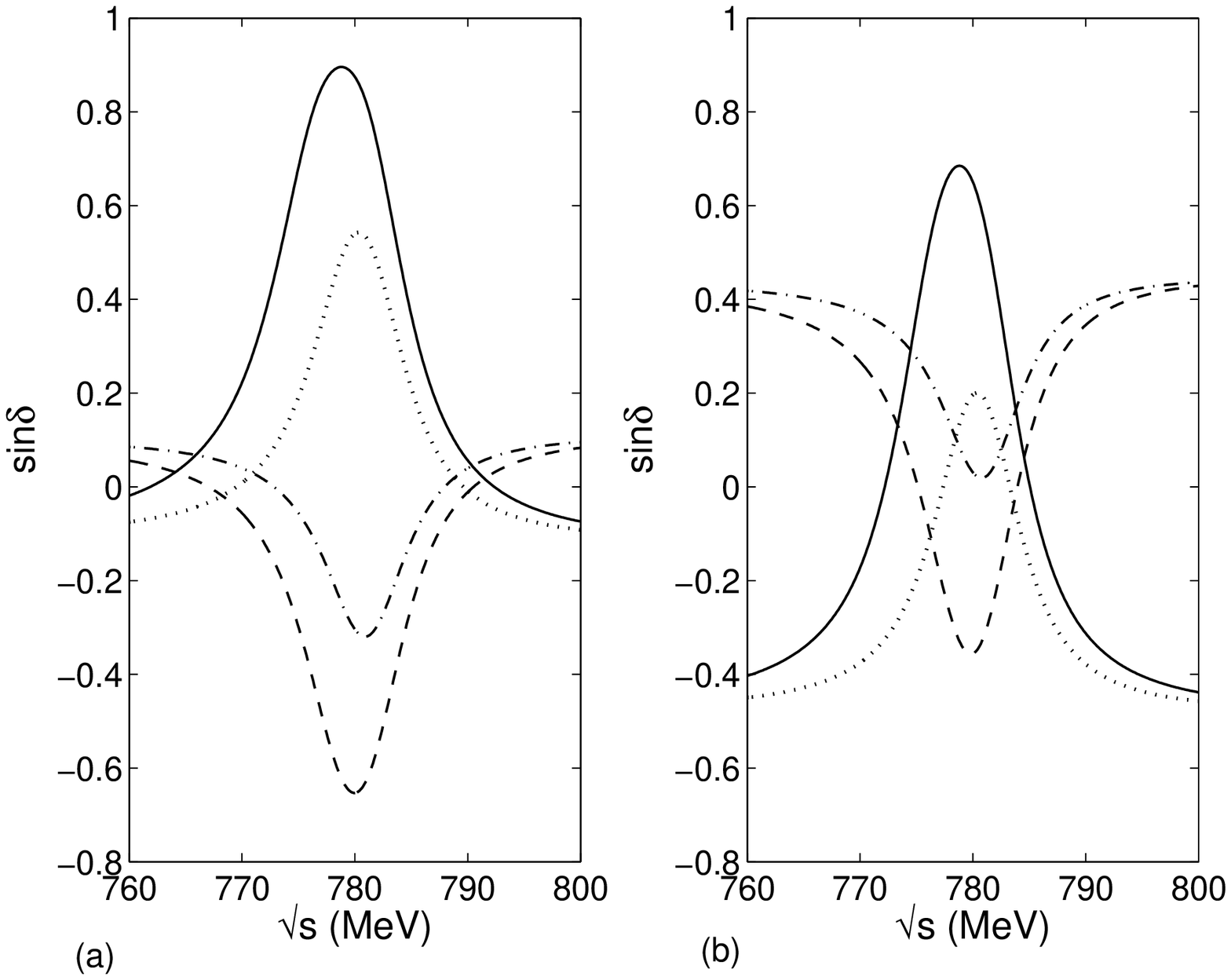}\\
\centerline{Fig. 3}
\end{figure}

\begin{figure}
\includegraphics[width=0.48\textwidth]{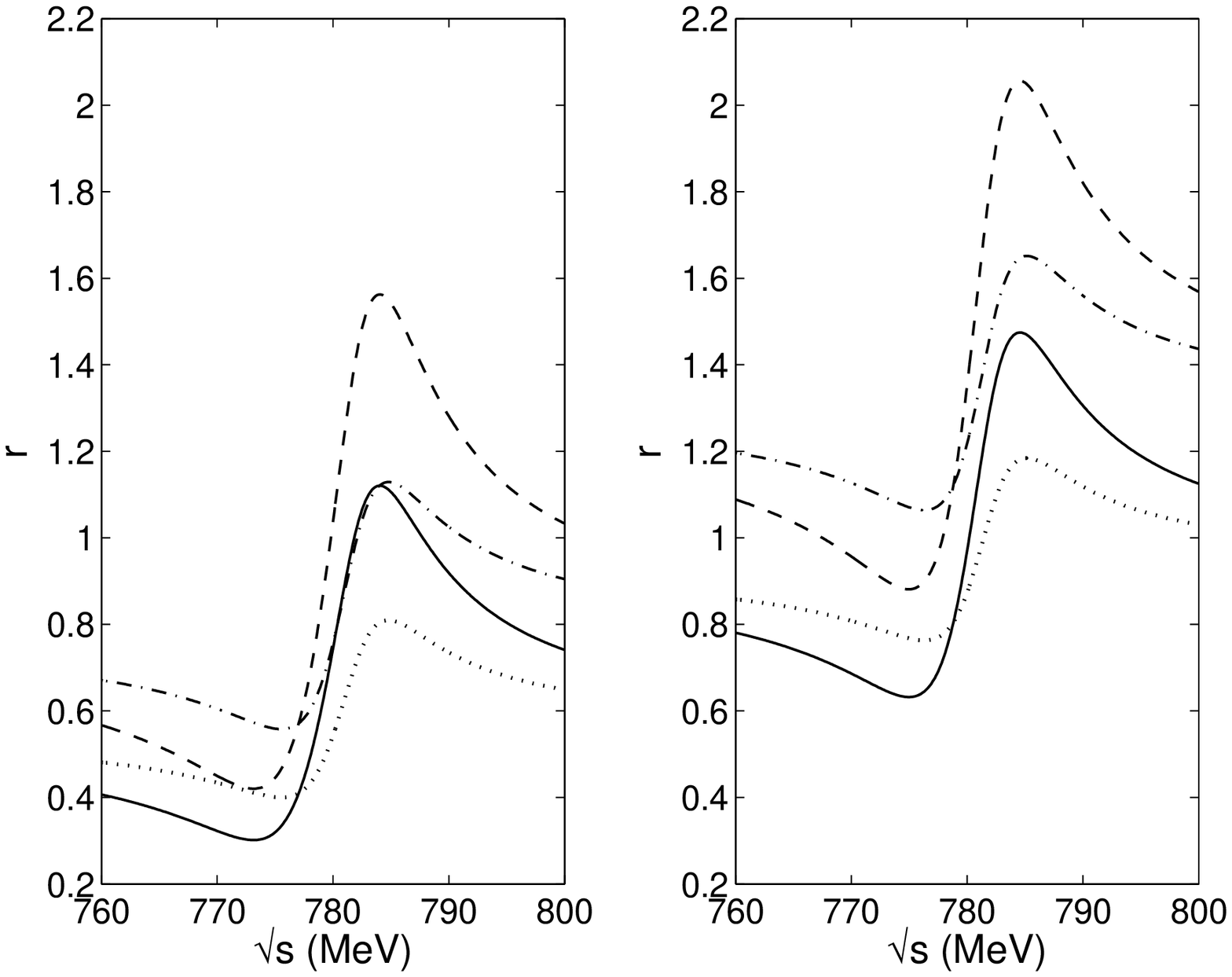}\\
\centerline{Fig. 4(a)}
\end{figure}
\begin{figure}
\includegraphics[width=0.48\textwidth]{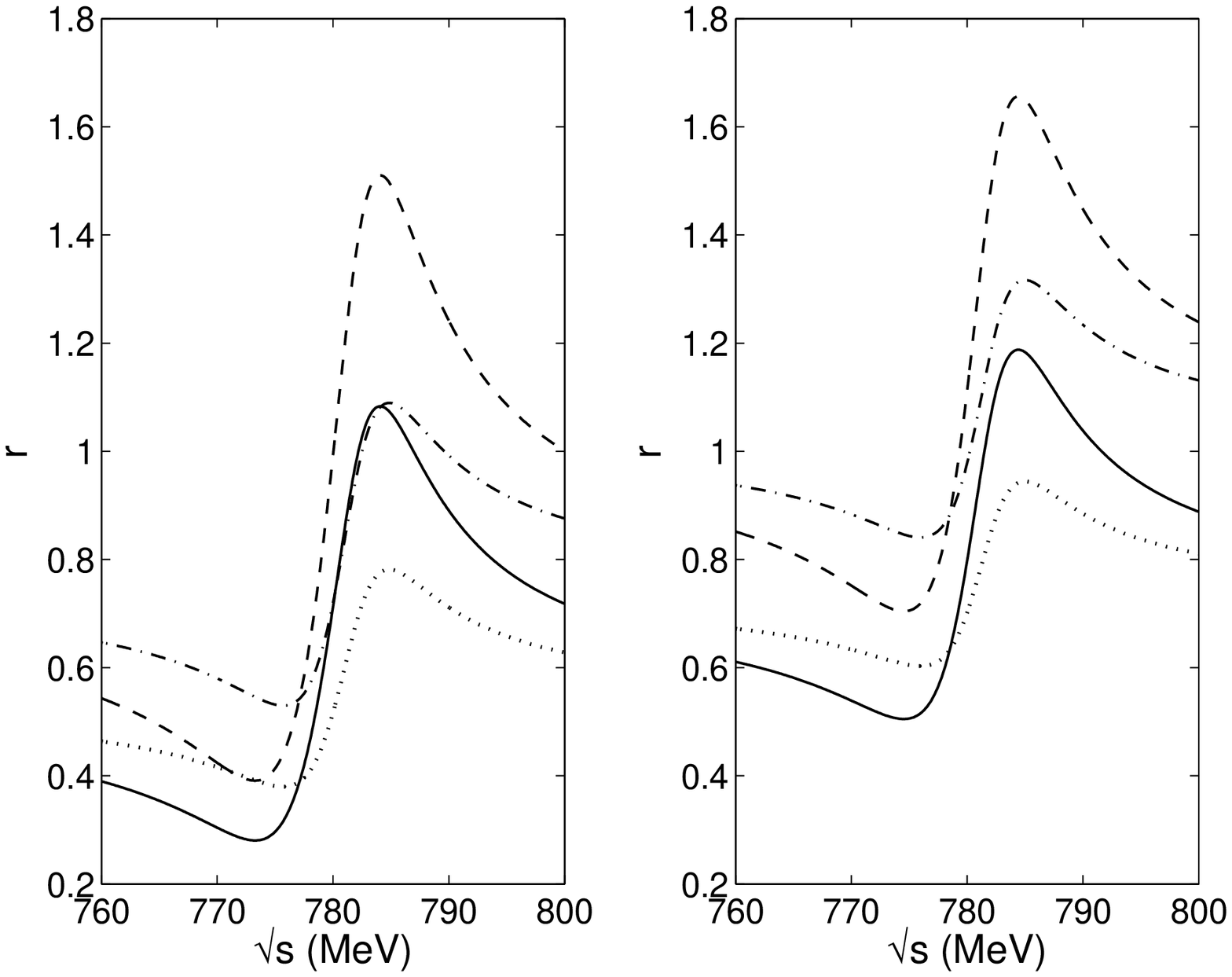}\\
\centerline{Fig. 4(b)}
\end{figure}

\begin{figure}
\includegraphics[width=0.48\textwidth]{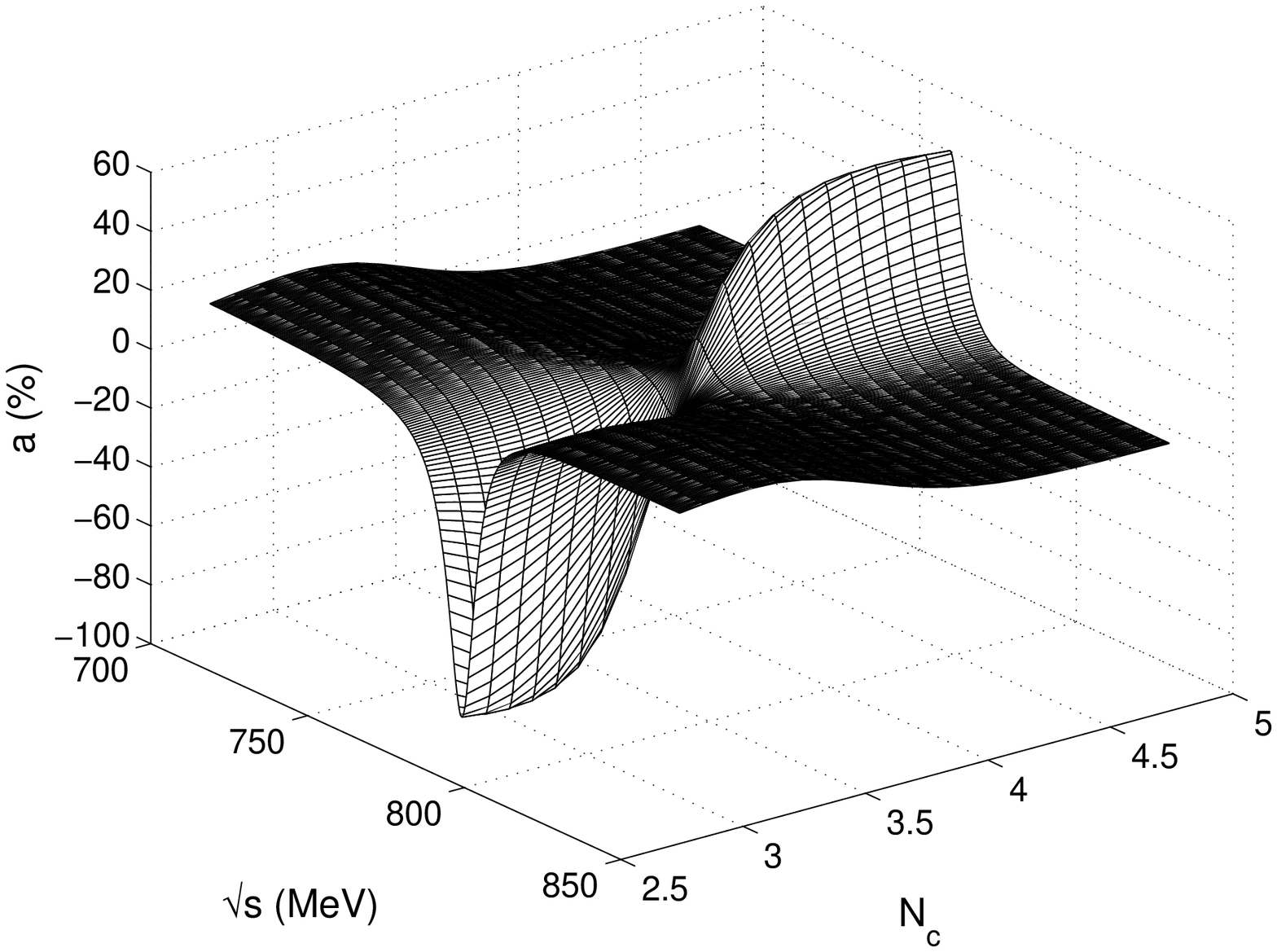}\\
\label{fig8} \centerline{Fig. 5(a)}
\end{figure}
\begin{figure}
\includegraphics[width=0.48\textwidth]{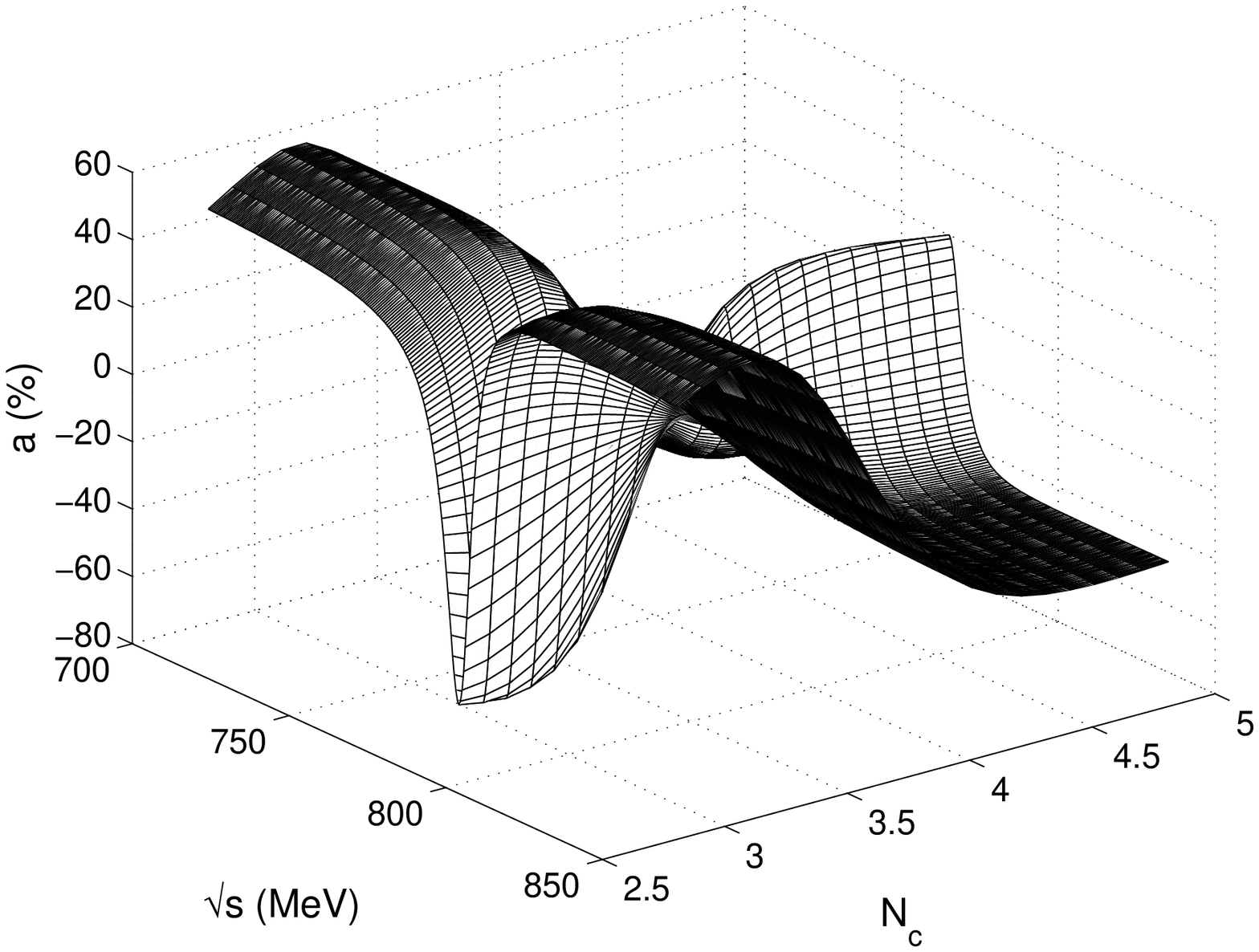}\\
\label{fig9} \centerline{Fig. 5(b)}
\end{figure}

\begin{figure}
\includegraphics[width=0.48\textwidth]{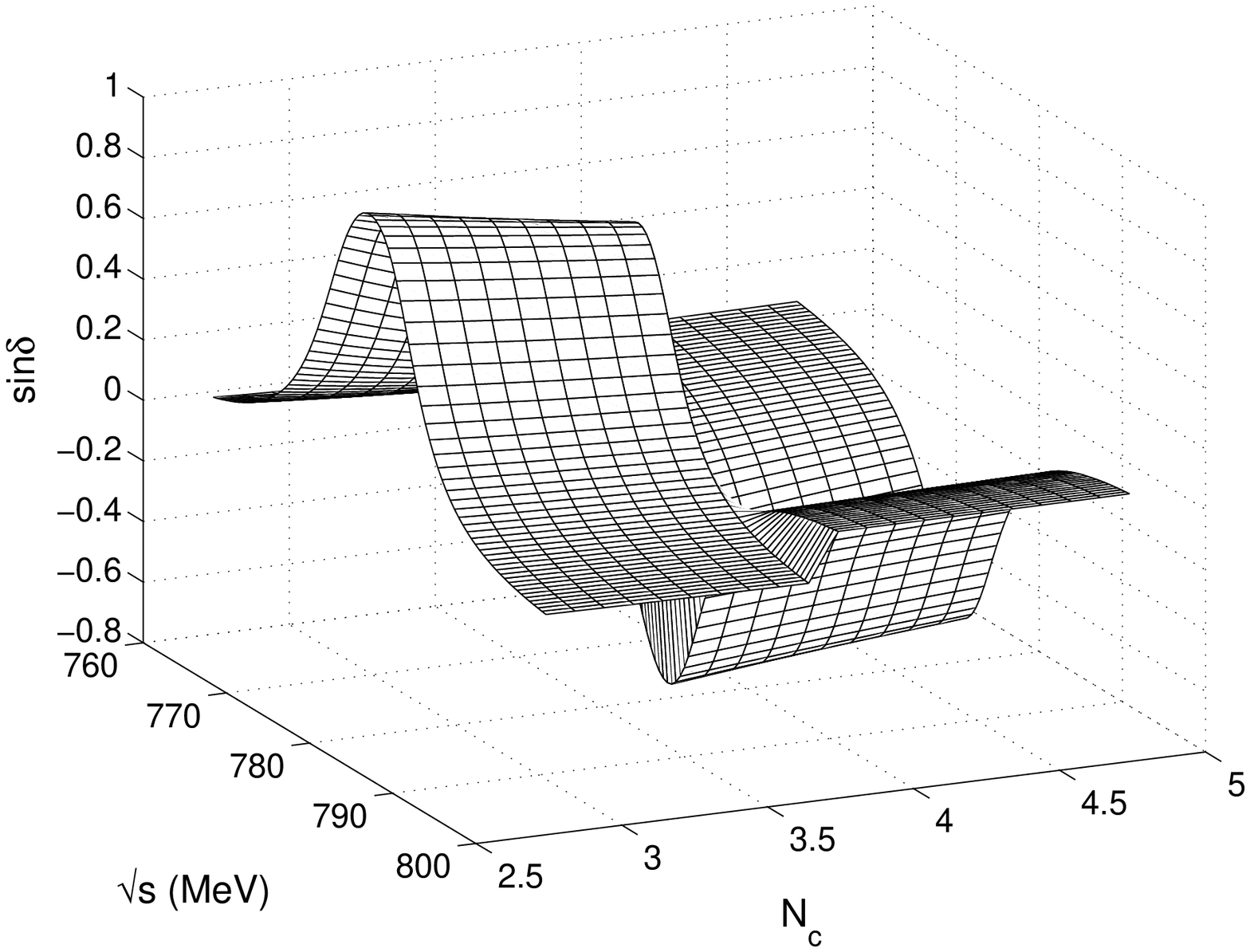}\\
\label{fig6} \centerline{Fig. 6(a)}
\end{figure}

\begin{figure}
\includegraphics[width=0.48\textwidth]{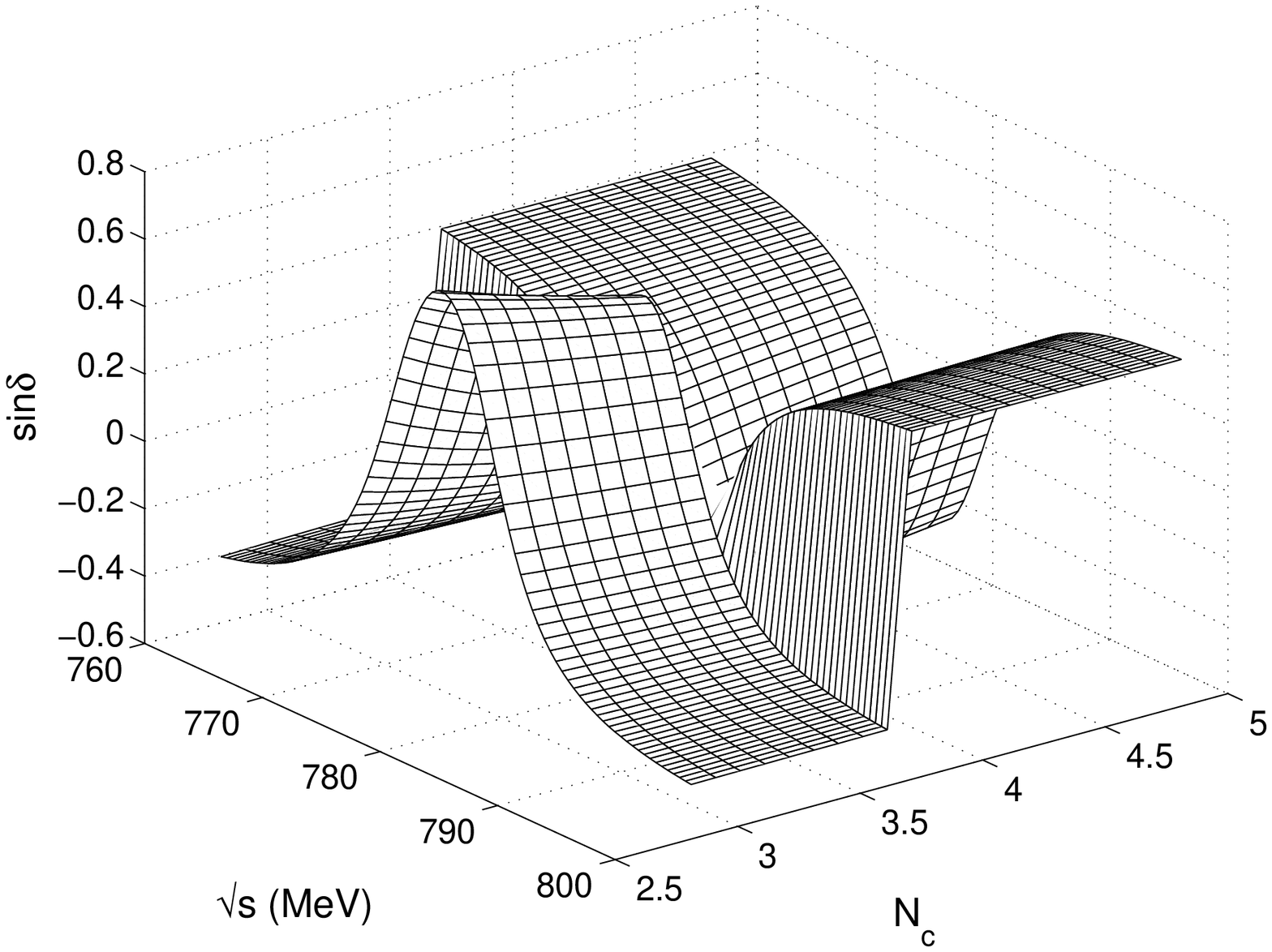}\\
\label{fig7} \centerline{Fig. 6(b)}
\end{figure}

\begin{figure}
\includegraphics[width=0.48\textwidth]{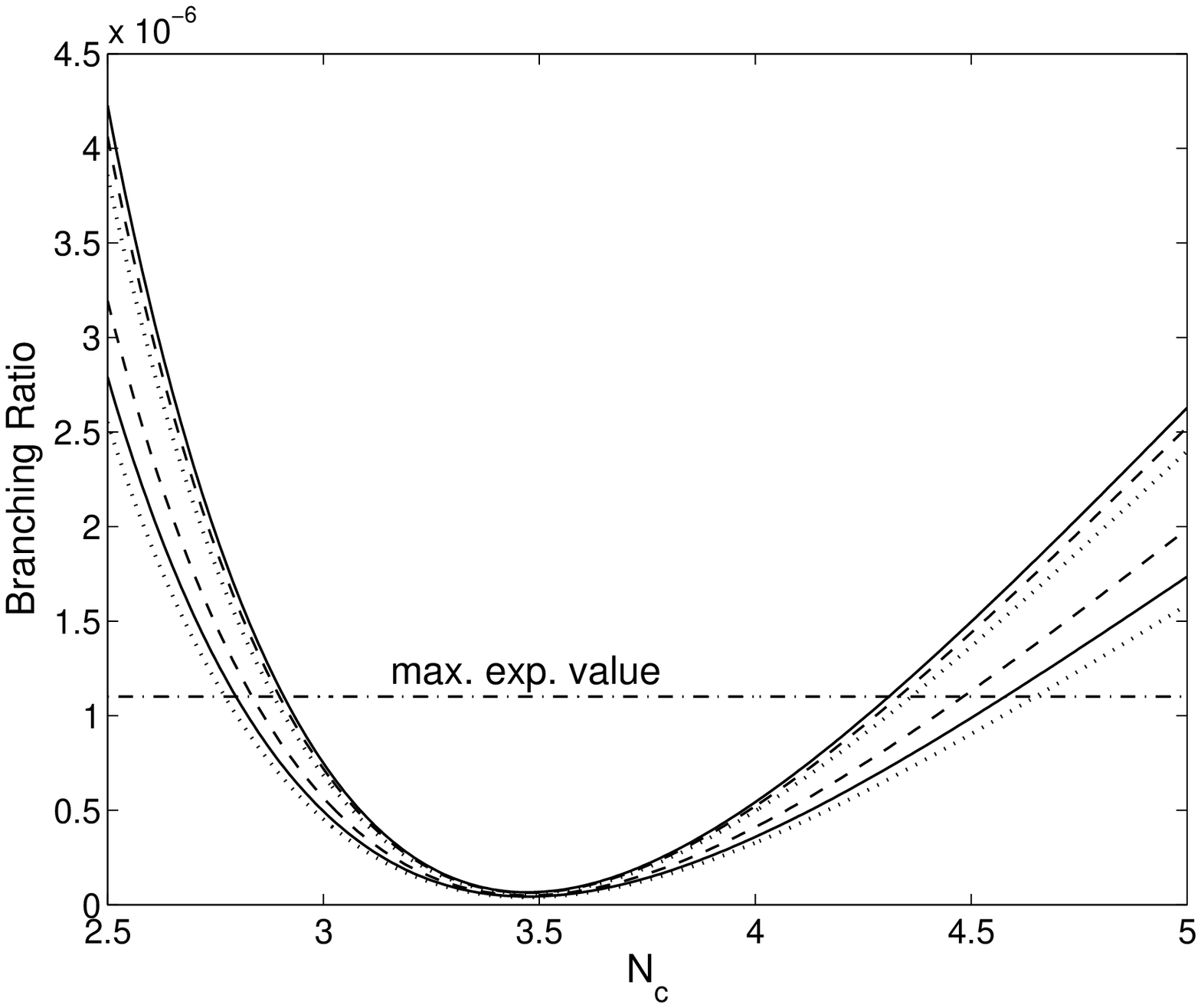}\\
\label{fig10} \centerline{Fig. 7}
\end{figure}

\begin{figure}
\includegraphics[width=0.48\textwidth]{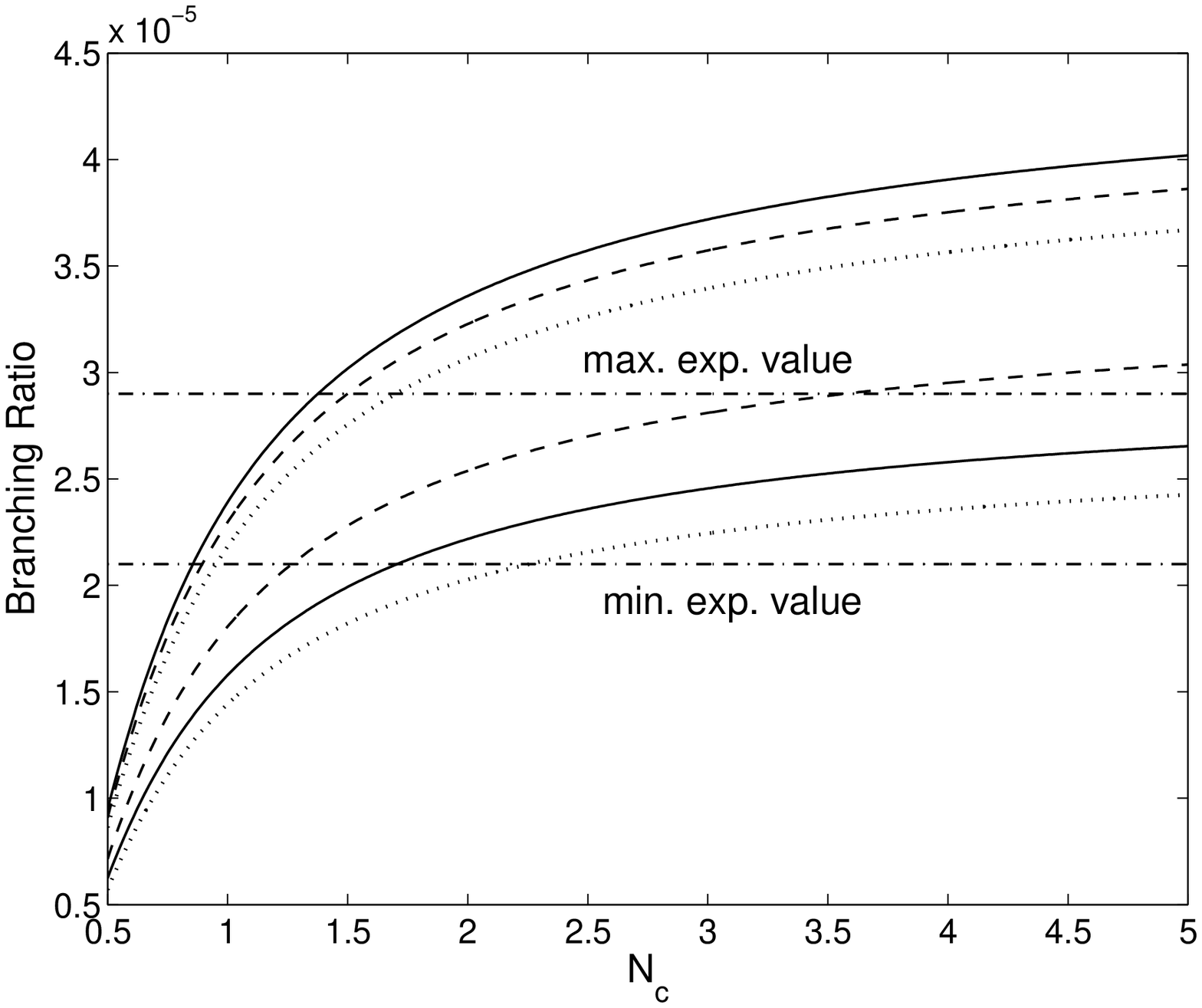}\\
\label{fig11} \centerline{Fig. 8}
\end{figure}

\end{document}